\documentclass[pre,twocolumn,showpacs,superscriptaddress]{revtex4}
\usepackage{amsmath,amssymb}
\usepackage{graphics,color}
\usepackage{times}

\newcommand{\aver}[1]
{\left \langle #1 \right \rangle}

\renewcommand{\Re}{\textrm{Re}}

\begin{document}

\title{Avoided level crossing statistics in open chaotic billiards}

\author{Charles Poli}
\affiliation{Laboratoire de Physique de la Mati\`ere Condens\'ee, CNRS UMR 6622
\\Universit\'e de Nice-Sophia Antipolis - 06108 Nice cedex 2, France}
\author{Barbara Dietz}
\affiliation{Institut f\"ur Kernphysik, Technische Universit\"at Darmstadt, Schlossgartenstrasse 9, 64289 Darmstadt, Germany}
\author{Olivier Legrand}
\author{Fabrice Mortessagne}
\affiliation{Laboratoire de Physique de la Mati\`ere Condens\'ee, CNRS UMR 6622
\\Universit\'e de Nice-Sophia Antipolis - 06108 Nice cedex 2, France}
\author{Achim Richter}
\affiliation{Institut f\"ur Kernphysik, Technische Universit\"at Darmstadt, Schlossgartenstrasse 9, 64289 Darmstadt, Germany}
\affiliation{ECT$^*$, Villa Tambosi, I-38100 Villazzano (Trento), Italy}
\date{\today}

\begin{abstract}

We investigate a two-level model with a large number of open decay channels in order to describe avoided level crossing statistics in open chaotic billiards. This model allows us to describe the fundamental changes of the probability distribution of the avoided level crossings compared with the closed case. Explicit expressions are derived for systems with preserved and broken Time Reversal Symmetry (TRS). We find that the decay process induces a modification at small spacings of the probability distribution of the avoided level crossings due to an attraction of the resonances. The theoretical predictions are in complete agreement with the recent experimental results of Dietz \textit{et al.} (Phys. Rev. E {\bf 73} (2006) 035201).

\end{abstract}

\maketitle
It is by now established, that classical chaos manifests itself in universal spectral 
fluctuation properties of the eigenvalues of the corresponding quantum system.
They coincide with those of random matrices from the Gaussian Orthogonal Ensemble (GOE), if Time Reversal Symmetry TRS holds, 
from the Gaussian Unitary Ensemble (GUE), if TRS is broken \cite{Boh84,Meh91, Boh95, Guh98}.

Investigations of the universality of spectral fluctuation properties of classically chaotic systems range from nuclear physics \cite{Gar64,Wei09}, to systems  in other areas, like microwave billiards \cite{Gra92,So95,Sto02,Bar05}, optical experiments \cite{Gma98,Mic07}, quantum dots \cite{Mar92,Alh00}, and acoustic setups \cite{Wea89,Ell95,Ber99}.  In systems depending on a global parameter, the correlations between eigenvalues at different parameter  values show a universal behavior, which again is well described by random matrix theory (RMT) \cite{Zak91,Zak93a,Zak93b,Leb99}. In some cases, implying a local parameter, RMT fails, as reported in reference \cite{Bar99}.

In an experiment presented in \cite{Die06} the spectral properties of a superconducting microwave billiard, whose boundary was varied parametrically, were investigated. It models a quantum billiard of corresponding shape, whose classical dynamics is chaotic. The observed deviations from the expected GOE behavior were attributed to the measurement process. Indeed, resonance spectra of a microwave billiard are measured by connecting it to the exterior via emitting and receiving antennas. Thus the resonator is an
open system with the antennas acting as single scattering channels.
The influence of the flux of microwave power flowing from the emitting to the
receiving antenna on the spectral properties of the system is so weak that it
cannot be detected through spectral measures like the nearest neighbor spacing distribution or the $\Sigma^2$ statistics at a fixed value of the parameter. The distribution of the avoided crossings of the eigenvalues as function of the parameter on the other
hand showed deviations from the GOE result, which were attributed to the openness of the
resonator. These assumptions were confirmed by numerical simulations based on a random matrix model for parameter dependent, chaotic and open
systems. The aim of the present paper is the derivation of an analytic expression for the avoided-crossings distribution of such systems. It goes in line with that for the corresponding distribution for closed systems with and without TRS, which is based on an ensemble of two-dimensional random matrices \cite{Zak91}.
 
We develop our approach within the framework of an effective Hamlitonian model \cite{Oko03}. To describe statistical properties of avoided crossings in open systems, we introduce the effective Hamiltonian $H_{eff}$ which depends on a continuous parameter $\mu$ through its Hermitian part \cite{Die06}     
\begin{equation}\label{heff}
H_{eff}(\mu) =H(\mu)-\frac{i}{2}VV^T\, .
\end{equation}
Here $H(\mu)$ is the Hamiltonian of the closed system modeled by a $2\times 2$  random matrix and $iVV^T/2$ is an imaginary potential describing the coupling to the environment in terms of $M$ open channels. The $2 \times M$ matrix $V$ contains the coupling amplitudes $V_n^m$ which couple the $n$th level to the $m$th open channel. As a result, the eigenvalues of the effective Hamiltonian are complex, $\epsilon_\pm=E_\pm-\frac{i}{2}\Gamma_\pm$, where $E_\pm$ and $\Gamma_\pm$ are, respectively, the two eigenenergies and  the two spectral widths of the 2-level model. For the study of statistical properties $H$ is replaced by a Gaussian random matrix \cite{Sto99} and the matrix elements $V_{n}^m$ are chosen to be Gaussian-distributed with zero mean and variance $\sigma^2=2\lambda\Delta$, where $\lambda$ is the coupling strength and $\Delta$ is the mean level spacing of the closed system \cite{Sav06}. In the eigenbasis of $H(\mu)$ (the $\mu$-dependence is omitted in the following), the effective Hamiltonian is written as
\begin{equation}\label{mod}
H_{eff} =\begin{pmatrix} 
E_1 &  & 0\\
0 &  & E_2 
\end{pmatrix} 
-\frac{i}{2} \begin{pmatrix} 
\Gamma_{11} &\Gamma_{12} \\
\Gamma_{21} & \Gamma_{22}   \\
\end{pmatrix}\, ,
\end{equation}
where $E_{1,2}$ are the $\mu$-dependent eigenenergies of $H$ ($E_2>E_1$ is assumed) and $\Gamma_{np}=\sum_{c=1}^M V_n^mV_p^m$. Note that the model is applicable only as long as the coupling is weak enough so that the spectral widths remain of the same order of magnitude \cite{Haa92,Leh95}. The complex eigenvalues of the effective Hamiltonian $H_{eff}$ in Eq. (\ref{mod}) read
\begin{equation}\label{epm}
\epsilon_{\pm}=\frac{E_1+E_2-\frac{i}{2}(\Gamma_{11}+\Gamma_{22}) \pm \sqrt{D}}{2}
\end{equation}
with
\begin{equation}\label{delta}
D=\Big ((E_1-E_2)+\frac{i}{2}(\Gamma_{22}-\Gamma_{11})\Big )^2-\Gamma_{12}\Gamma_{21}\, .
\end{equation}
The spacing between the two eigenergies, $d= \epsilon_{+}-\epsilon_{-}$ can be read off from (\ref{epm}) 
and (\ref{delta}), 
\begin{equation} \label{dD} 
d=\Re(\sqrt{D})\, .
\end{equation}
Considering the limit of a large number of open channels $M$ in the weak coupling regime, we may apply the central limit theorem and replace the random variables depending on the coupling amplitudes by their averages,
\begin{equation}
\aver{\Gamma_{nn}}=M\sigma^2 \qquad \qquad \aver{\Gamma_{np}\Gamma_{pn}}=M\sigma^4\, .
\end{equation}
Then the spacing $d$ is given by 
\begin{equation}\label{d} 
d=
    \begin{cases}
        \sqrt{s^2-M\sigma^4}& \text{if } s>\sqrt{M}\sigma^2 \\
        0 & \text{otherwise}
    \end{cases}\ ,
\end{equation}
where $s=E_2-E_1$ is the spacing of the eigenenergies of the closed system. Note that $M\sigma^4=\text{var}(\Gamma)/2$ implies that the modifications on the spacings due to the openness of the system are related to the fluctuations of the spectral widths \cite{Pol09}. 
In the limit $M\rightarrow\infty$ and $\sigma^2\rightarrow 0$ with $M\sigma^2=\aver{\Gamma}$ fixed, $\text{var}(\Gamma)\rightarrow 0$ and thus the spacing between eigenenergies of the open system converges to that of the closed system, $d \rightarrow s$, in spite of non-vanishing losses. One of the effects of the imaginary potential is that the eigenvalues mutually attract each other along the real axis \cite{Mul95}. As this attraction increases when $s$ decreases, the local minima of both $d$ and $s$ coincide. In other words, the values of the parameter $\mu$ at the avoided crossings are the same for the closed and the open system. Accordingly, in the derivation of the distribution of avoided crossings $c$ of the open system, the spacings $s$ are assumed to be distributed as the avoided crossings of the corresponding closed system. With Eq.~(\ref{d}) the probability distribution of the avoided level crossings $p(c)$ is given by
\begin{multline}\label{pcdef}
p(c)=\aver{\delta(c)\theta(\sqrt{M}\sigma^2-s)}  \\
+\aver{\delta\Big(c-\sqrt{s^2-M\sigma^4}\Big)\theta(s-\sqrt{M}\sigma^2)}\, ,
\end{multline}
where $\theta$ is the Heaviside step function and the triangular brackets denote averaging with respect to the spacing $s$.

For closed chaotic systems with TRS the probability distribution of avoided crossings has been calculated by Zakrzewski and Ku\ifmmode  \acute{s}\else \'{s}\fi{} \cite{Zak91},
\begin{equation}\label{psGOE}
p(s)=\sqrt{\frac{2}{\pi\alpha^2}}e^{-s^2/(2\alpha^2)}\, ,
\end{equation}
where the mean value of $s$ is given by $\aver{s}=\alpha\sqrt{2/\pi}$.  Averaging over $s$ yields 
\begin{multline}\label{pcGOE}
p(c)=\text{erf}\Big(\frac{\sqrt{M}\sigma^2}{\sqrt{\pi}\alpha}\Big)\delta(c) \\
+\sqrt{\frac{2}{\pi\alpha^2}}\frac{c\; e^{-(c^2+M\sigma^4)/(2\alpha^2)}}{\sqrt{c^2+M\sigma^4}}\, ,
\end{multline}
where $\alpha$ fixes the average of $c$. Note that the behavior of $p(c)$ at small spacings differs strongly from the GOE prediction (\ref{psGOE}). The linear behavior of the  distribution induces a dip and the local minima of the spacings have a zero-crossings contribution leading to the presence of a $\delta$-peak at the origin. This peak is neither restricted to 2-level models nor to a large number of channels. It was also found numerically in \cite{Die06} where an effective Hamiltonian with 1000 levels and $M=3$ open channels was considered and is characteristic of non-Hermitian Hamiltonians of the form Eq.~(\ref{heff}) \cite{Dem01,Hei98,Hei00,Kec03}.

In the experimental setup \cite{Die06}, only finite spacings could be measured due to the discrete sampling of the data. Therefore, to compare theory and experiment it is more convenient to consider the distribution of non zero avoided crossings $c'$,
\begin{equation}\label{pcnGOE}
p(c')=\sqrt{\frac{2}{\pi\alpha^2}}\frac{c'\;e^{-(c'^2+M\sigma^4)/(2\alpha^2)}}{\text{erfc}\Big(\frac{\sqrt{M}\sigma^2}{\sqrt{\pi}\alpha}\Big)\sqrt{c'^2+M\sigma^4}}\, .
\end{equation} 

The analysis can be extended to open, parameter dependent, chaotic systems with broken TRS. For closed systems of this type the probability distribution of avoided crossings \cite{Zak91} reads
\begin{equation}\label{psGUE}
p(s)=\frac{s}{2\alpha^2}e^{-s^2/(4\alpha^2)}\, ,
\end{equation}
Using the same effective Hamiltonian model (\ref{mod}) i.e. considering real coupling amplitudes, the probability distribution of avoided level crossings is derived using (\ref{pcdef}) and (\ref{psGUE}),
\begin{equation}\label{pcGUE}
p(c)=(1-e^{-M\sigma^4/(4\alpha^2)})\delta(c)+\frac{c}{2\alpha^2}e^{-(c^2+M\sigma^4)/(4\alpha^2)}\, .
\end{equation}
Again a $\delta$-peak appears at $c=0$ due to the attraction of the eigenvalues on the real axis. However, in contrast to the GOE case, the probability distribution of the non zero avoided level crossings coincides with that of  the closed system given in Eq.~(\ref{psGUE})
\begin{equation}\label{pcnGUE}
p(c')=\frac{c'}{2\alpha^2}e^{-c'^2/(4\alpha^2)}\, .
\end{equation}
This robustness of GUE was previously observed in room temperature microwave billiards with broken TRS \cite{Sto95}.

\begin{figure}[ht]
\begin{center}
\scalebox{0.16}{\includegraphics{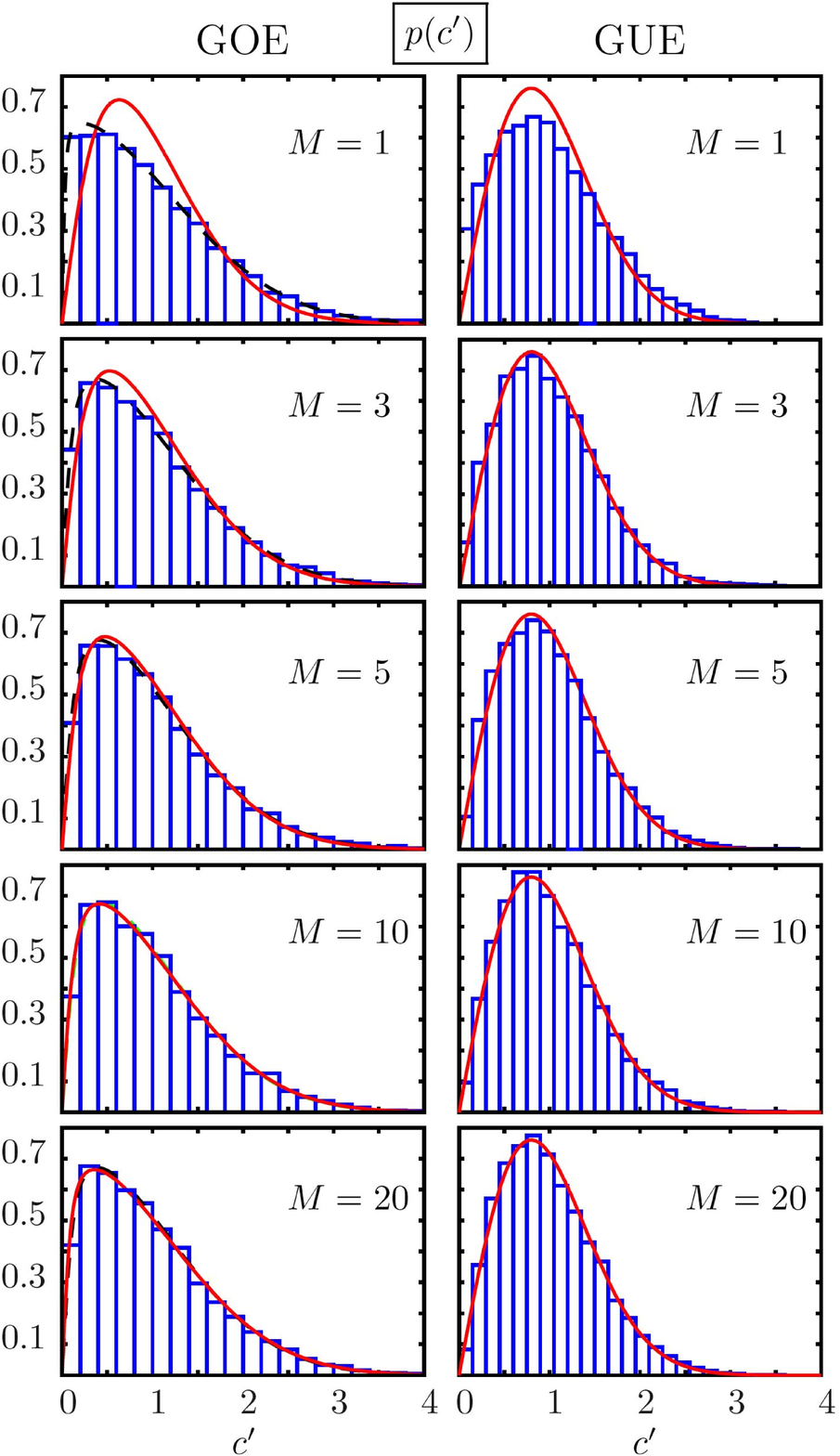}}
\end{center}
\caption{(color online) Probability distributions of non zero avoided level crossings for GOE  (left side) and for GUE (right side).  The number of open channels is $M$=1, 3, 5, 10, 20 and the coupling strength $\lambda=\sigma^2/(2\Delta)$ equals $\lambda=0.250, 0.083, 0.05, 0.025, 0.013$ from top to bottom. The histograms show the numerical simulations, the analytic distributions are shown as straight lines.  The dashed curves result from a fit of Eq.~(\ref{pcnGOE}) with $\lambda$ as a parameter to the numerical distributions, resulting in the effective coupling strengths $\lambda_{eff}=0.020, 0.038, 0.036, 0.028, 0.016$. For all curves, $\alpha$ is chosen such that $\aver{c'}=1$.}
\label{pcnfig}
\end{figure}

To analyze the evolution of both distributions for a small or not too large number of channels, numerical random matrix simulations were performed, where the eigenvalues of the closed, parameter dependent system were chosen as the  eigenvalues of the random matrix  
\begin{equation}\label{hmu}
H(\mu)=H_1\cos \mu+H_2\sin \mu\, ,
\end{equation}
simulating the closed system (see \cite{Die06}). Here $H_1$ and $H_2$ belong to the GOE or the GUE for the simulation of systems with or without TRS, respectively, and the coupling amplitudes $V_n^m$ with $n=1,\cdots , N$ and $m=1, \cdots , M$ are random Gaussian variables. Note that this model ensures that the mean level spacing is independent of $\mu$ \cite{Zak93b}. In the simulations, the matrices $H_1$ and $H_2$ are of size $1000\times 1000$, the variances of their elements are chosen equal and such that $\Delta=1/1000$, the parameter $\mu \in [0,\pi [$ is discretized in steps of $\delta \mu=\pi/1300$.  To ensure a fairly constant mean level spacing only the 400 resonances at the center of the Wigner semicircle were kept. To mimic the experimental resolution, a cut-off $c_0=0.1\Delta$ is introduced such that only values of $c$ larger than $c_0$ are used to build the numerical distributions. Furthermore the average of the spectral widths is fixed to $\aver{\Gamma} =0.5\Delta$, well away from the strong coupling regime \cite{Die06}. A comparison between analytic and numerical results is presented in Fig.\ref{pcnfig}.

For the GOE case a good agreement between the numerical and analytical descriptions is found for $M \ge 5$. For a smaller number of channels, the histograms are reproduced by choosing the coupling strength $\lambda=\sigma^2/(2\Delta)$ in Eq.~(\ref{pcnGOE}) as a parameter to obtain an effective coupling strength $\lambda_{eff}$ by means of a fit based on a least square algorithm. Thus, it appears that the expression (\ref{pcnGOE}), derived using the central limit theorem, can be extended to any $M$, considering $\lambda$ as a free parameter. 

The right column of Fig.~\ref{pcnfig} shows the results obtained for the GUE case. The prediction is in excellent agreement with the numerical results except for the case $M=1$. This is due to the small number of events at small distances of $p(s)$. Indeed, while important changes appear for the GOE due to the large number of small avoided crossings, the GUE case is only slightly modified because of a vanishing density of avoided crossings at the origin for the closed case. Note that  the distribution (\ref{pcnGUE}) is independent of the coupling strength such that a fitting procedure is not possible.

\begin{figure}[ht]
\begin{center}
\scalebox{0.17}{\includegraphics{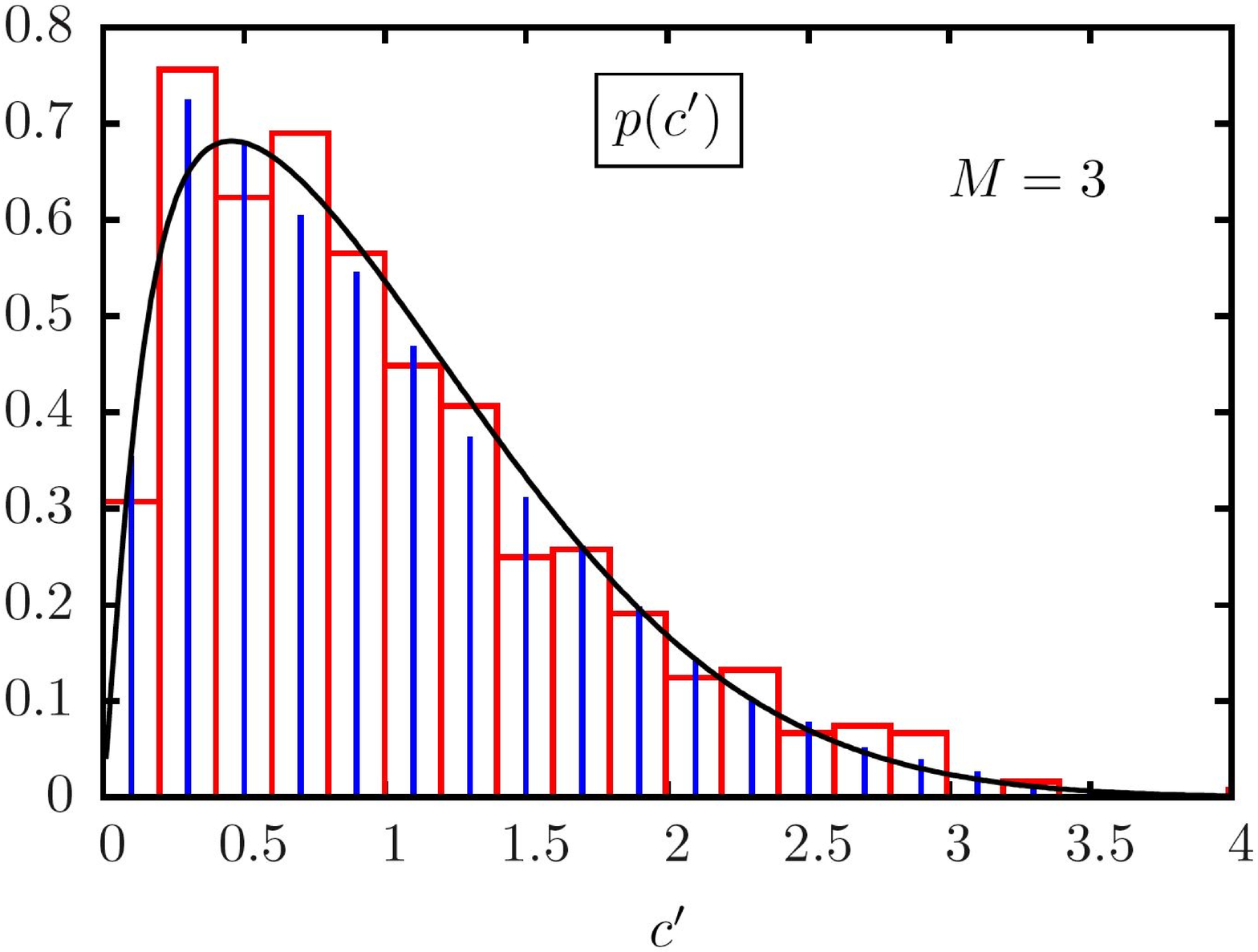}}%
\end{center} 
\caption{(color online) In boxes the experimental distribution of avoided crossings  \cite{Die06}. The continuous line shows the analytical prediction with $\lambda$=0.058 (obtained through a least square procedure). The vertical bars represent the numerically obtained distribution with $M=3$ and $\lambda$=0.02 \cite{Die06}. For all curves, the average is chosen such that $\aver{c'}=1$.}
\label{pcExpTheFig}
\end{figure} 

Now, let us finally compare the analytical prediction (\ref{pcnGOE}) with the experimental results of \cite{Die06} obtained using a superconducting microwave cavity, thus minimizing dissipative processes. Three antennas were attached to the cavity: they correspond, in our model, to $M=3$ open channels \cite{Alt95}. Absorption into the walls could be mimicked by additional fictitious weakly coupled channels \cite{Bar05,Sav06,Bro97}, however, its influence can be safely neglected in the analysis of the experimental data. Due to the lack of an analytic expression for the distribution of avoided crossings in open systems, the experimental distribution was compared with numerical simulations based on an effective Hamiltonian of the form Eq.~(\ref{heff}) with the parameter dependent Hamiltonian of the closed system given in Eq.~(\ref{hmu}). In reference \cite{Die06}, a good agreement beween both distributions was shown for values of $\lambda\simeq 0.02$. Note that due to the lack of an analytical expression this value was not determined from a fit and thus is only vague. A comparison between the experimental (boxes) and numerical histograms (vertical bars) is shown in Fig.~\ref{pcExpTheFig}. The numerical distribution has been computed with $\lambda =0.02$. The analytic result obtained through a fit using $\lambda$ as a parameter (continuous curve) is also shown in Fig.~\ref{pcExpTheFig}. The 2-level model result follows closely the experimental histogram. This confirms the interpretation drawn in \cite{Die06} that the deviation of the avoided-crossings distribution from the predicted GOE result for closed systems is due to the measurement process, \textit{i.e.} the influence of the three antennas, which couple the resonator modes inside the resonator to the exterior.

In summary, we have derived an analytic expression for the distribution of the avoided crossings of the resonances of quantum chaotic open systems based on a simple two-level random matrix model. Analytical results prove that the openness essentially modifies the avoided crossing distributions at small spacings. The theoretical predictions are in excellent agreement with numerical random matrix simulations for a number of open channel $M\ge 5$ in the GOE case and for $M \ge 3$ for GUE systems. For systems preserving TRS with a small number of open channels good agreement is achieved by using the coupling strength $\lambda$ as a fit parameter.

Finally, let us mention that the 2-level model can also be used to calculate the nearest level spacing distribution (NLSD) for open chaotic systems with a large number of channels. Whereas, for the strong coupling regime, the NLSD is substantially modified by the introduction of dissipation \cite{Haa96}, in the weak coupling case, due to a vanishing density at small spacings for both GOE and GUE, the NLSD for open systems will only be moderately modified, as pointed out above.

\begin{acknowledgments}
We would like to thank Patricio Leb{\oe}uf for fruitful discussions and Ulrich Kuhl for a critical reading of the manuscript. B.~D.\ and A.~R.\ are grateful for the financial support by the DFG within SFB~634.
\end{acknowledgments}


\end{document}